# Ledgerdata Refiner: A Powerful Ledger Data Query Platform for Hyperledger Fabric

Ence Zhou
*Information Technology Laboratory*
*Fujitsu Research & Development Center*
Suzhou, China
zhouence@cn.fujitsu.com

Haoli Sun
*Information Technology Laboratory*
*Fujitsu Research & Development Center*
Suzhou, China
sunhaoli@cn.fujitsu.com

Bingfeng Pi
*Information Technology Laboratory*
*Fujitsu Research & Development Center*
Suzhou, China
winter.pi@cn.fujitsu.com

Jun Sun
*Information Technology Laboratory*
*Fujitsu Research & Development Center*
Beijing, China
sunjun@cn.fujitsu.com

Kazuhiro Yamashita
*Information Systems Technologies Laboratory*
*Fujitsu Laboratories Ltd.*
Kawasaki, Japan
y-kazuhiro@jp.fujitsu.com

Yoshihide Nomura
*Information Systems Technologies Laboratory*
*Fujitsu Laboratories Ltd.*
Kawasaki, Japan
y.nomura@jp.fujitsu.com

***Abstract*—Blockchain is one of the most popular distributed ledger technologies. It can solve the trust issue among enterprises. Hyperledger Fabric is a permissioned blockchain aiming at enterprise-grade business applications. However, compared to traditional distributed database solutions, one issue of blockchain based application development is the limited data access. For Fabric, the ledger data can only be retrieved by limited interfaces provided by Fabric SDKs or chaincode. In order to meet the requirements of data query and provide flexible query functions for real applications built on Fabric, this paper proposed a ledger data query platform called Ledgerdata Refiner. With ledger data analysis middleware, we provide sufficient interfaces for users to retrieve block or transaction efficiently. It is also able to track historical operations for any specific state. In addition, schemas of ledger state have been analyzed and clustered, which enable users to perform rich queries against ledger data. Finally, we validate the effectiveness of our query platform on a real application.**

## I. Introduction

Blockchain technology was first introduced in Bitcoin [1] which is a distributed cryptocurrency system. Since then, different domains have shown growing interests in blockchain. Meanwhile, blockchain has witnessed the growth of numerous user cases. A blockchain is a shared, distributed ledger that records transactions and is maintained by multiple nodes in the network where nodes do not trust each other. Each node holds an identical copy of the ledger which is usually represented as a chain of blocks, which are linked using cryptography [2] . Each block in the chain contains lots of transactions and a cryptographic hash of its immediate previous block, thereby guarantee the immutability of blockchain ledger. The structure of blocks in a peer's ledger is depicted in Figure 1.

A blockchain network can be either permissionless or permissioned. In a permissionless network or public network such as Bitcoin and Ethereum [3], anyone can join the network to submit transactions. Permissionless blockchain networks power up most of the market's digital currencies. They allow each end user to create a personal address and to interact with the network by submitting transactions, and adding entries to the ledger. Additionally, all parties have the choice of running a node on the system, or employing the mining protocols (e.g., POW [1] or POS [4] ) to help to verify the transactions. In a permissioned blockchain or private blockchain, end users are required to be authenticated before joining the network. Permissioned blockchains are preferred by centralized organizations, which leverage the power of the network for their own, internal business operations. Company consortiums are also likely to employ private blockchains to securely record transactions, and exchange information between one another.

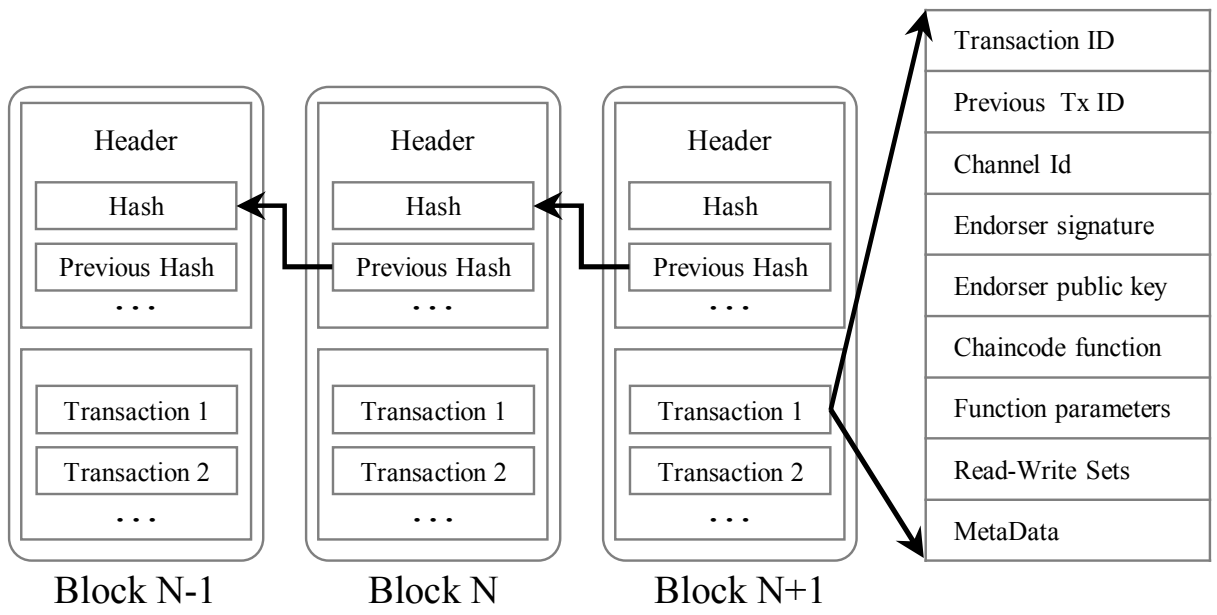

Fig. 1. The structure of blocks (take Hyperledger Fabric as an example)

In this paper, we focus on Hyperledger Fabric [5] that is one of the most popular permissioned blockchain platforms developed by the Linux foundation. Fabric is an open source permissioned blockchain platform which aims at enterprise-grade business applications. It has a modular design and a high performance which can reach about 1,000 transactions per seconds (TPS) [6] through trust models and pluggable components, while Ethereum is only around 15 TPS. More and more enterprises begin to build blockchain applications based on Fabric. Fabric is now there for used in many use cases such as Finance [7], Global Trade Digitization [8], IoT [9] and so on. It has gained a lot of attentions from both industry and academic in recent years [10, 11, 12, 13, 14]. Fabric is the focus of our data analysis study in this paper.

Fabric consists of various pluggable components such as endorsers, ordering service, a set of databases and committers. Due to numerous components and databases with different functions, Fabric provides various configurable parameters such as state database options and history index startup options. Since this paper focuses on data query in Fabric, one of the main challenges is how to analyze application data outside Fabric system. For example, depending on the application and requirements, one might need to answer the following questions:

- How many types of states have been stored in the world state database.
- If the history index is not open, how to get the operating history for a state.
- How to make rich query upon states when GoLevelDB [15] is adopted as state database.
- In addition to accessing data through chaincodes, is there any other ways to perform offline data analysis outside Fabric system?

To address the above challenges, we proposed a ledger data query and analysis framework which supporting real-time synchronization. This framework gets ledger data from a running blockchain network which builds on Hyperledger Fabric and stores the parsed results into a third-party database to provide a convenient way to browse ledger data. We perform a comprehensive empirical study of Fabric V1.1 [16], especially on the structures of blocks and transactions. Specifically, the three major contributions are listed below:

*1)* A convenient way to access different storage databases of Fabric in offline model by reorganising ledger data into a thrid-party database is proposed. There is no need to access ledger data through chaincodes.

*2)* Schema analysis of all stored states is proposed to let the user have a better overview of the ledger data. We also provide advanced query functions based on the analyzed state schemas.

*3)* A framework called Ledgerdata Refiner is proposed for detailed and real-time analyzing ledger data for Fabric system.

The rest of the paper is organized as follows. Section II surveys related work about data analysis on blockchain. Section III briefly describes transaction flow, blockchain structure and ledger data storages of Hyperledger Fabric. Section IV proposes the problems. Section □ presents our core contributions to data reorganization and the technical details of how to provide advanced query on ledger data. Section □ shows the experiments on a real application based on Fabric. Finally, we conclude this paper in Section □.

## II. Related Work

This paper focuses on the research of blockchain data analysis and query. In a permissionless network such as Bitcoin [1] and Ethereum [3], for their publicness, there are many institutions and scholars engaged in this research area. In order to prevent money laundering and to counter terrorism financing, researchers tried to reveal relationships among addresses by analyzing transactions and showed connectivity to graph structure for further understanding of Blockchain interactions [17, 18]. Bitcoin Block Explorer [19] is a web site provides friendly-UI to display detailed information about Bitcoin blocks, addresses, and transactions. For Ethereum data analysis and query, Chen et al. [20] proposed an approach to detect Ponzi schemes on blockchain by using data mining and machine learning methods on the transactions. Etherscan [21] is a tool similar to Block Explorer.

For permissioned blockchain networks (Hyperledger Fabric in particular), some research results have appeared. Ledgerfsck [22] was used to verify Hyperledger Fabric channel ledger. It traversed ledger starting from the very first block till the last one and utilized Message Crypto Service to verify blocks integrity. Hyperledger Explorer [23] was initially contributed by IBM, Intel and DTCC. And now it is a blockchain module and one of the Hyperledger projects hosted by the Linux Foundation. It is designed to create a user-friendly web application. Working with Hyperledger Fabric, it can view blocks, transactions and associated data, network information (name, status, list of nodes), as well as any other relevant information stored in the ledger. Splunk [24] built an agent to listen on peers for blocks, transactions and chaincode events and forwarded them to Splunk platform. Users could search all their data from infrastructure to ledgers in one place. Besides, some other tools in the same type, like Fabric Net Server [25], they have made some optimizations in data visualization. But they haven't improved much on ledger data analysis. So far all these tools don't support any complex query for blocks, transactions and the contents of states, they also cannot track a state's operating history.

It is evident from the related work that there lacks a platform or an approach to browse and query the ledger data of Fabric easily. That's the contribution of this paper. It contains three main functions: a) support for querying blocks and transactions in a variety of ways. b) For any state in ledger data, a function to track its operating history is proposed. c) And an overall view of state schemas for all the stored states and classify states by their schemas is presented. It supports rich queries for their contents, just like operating in a relational database.

## III. Basic Concepts

This section will introduce some basic conceptions in Hyperledger Fabric.

### *A. Transaction Flow in Hyperledger Fabric*

This paper focuses the ledger data analysis of Hyperledger Fabric. In order to have a better understanding of the data storages, we first introduce the complete process of a transaction, then show how the data storages supports the entire process at different stages. There are three main components, which are showed in Figure 2.

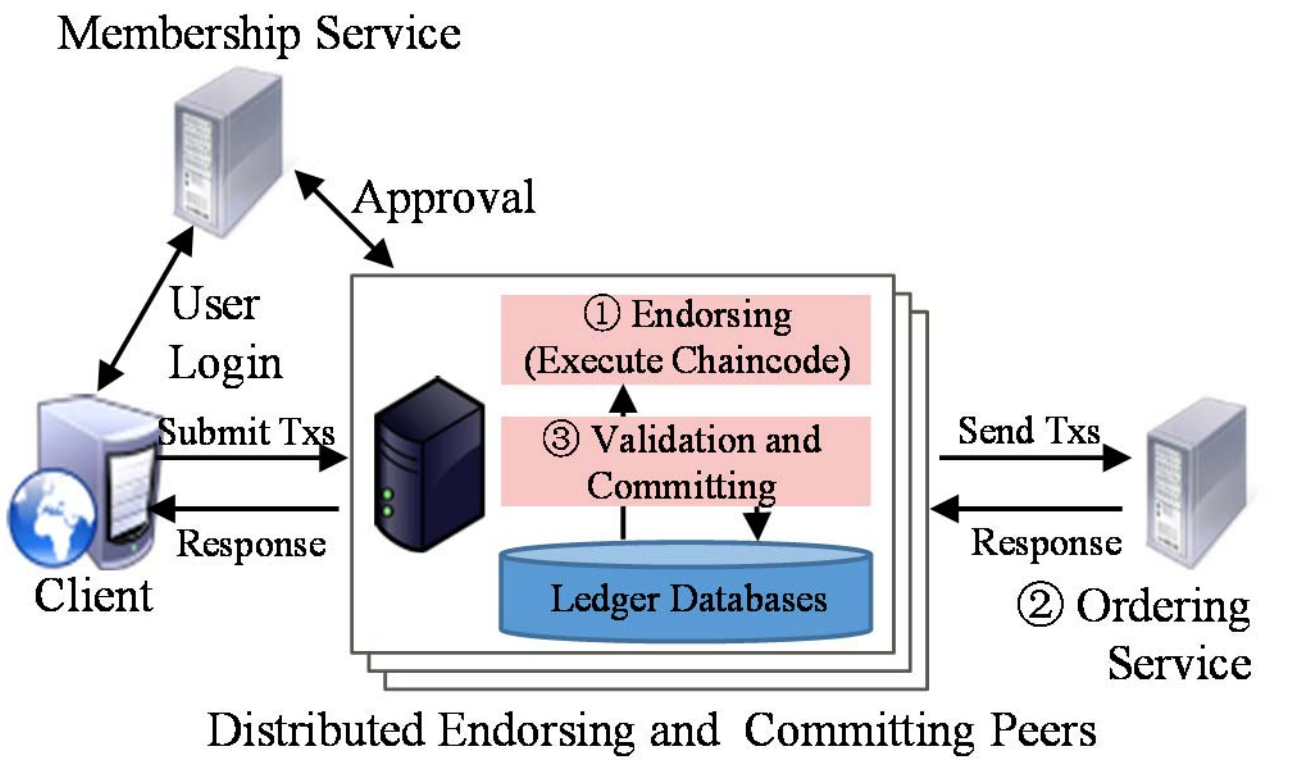

Fig. 2. Transaction Flow in Hyperledger Fabric

*1) Endorsement Phase:* A client sends its transaction to some endorsers. Each endorser executes the transaction in a sandbox and returns the endoring result to the client. Including signatures of the endorsing peers and the corresponding read-write set along with the version number of each state that was accessed. Then the client will send the transaction to the ordering service if the collected endorsements satisfy the endorsement policy [5]. In this stage, instantiated smart contract on each endorser will interact with the world state (a data storege in Fabric), which keeps the current value of ledger states.

*2) Ordering Service:* The ordering service receives transactions from different clients and sorts them by channels. It does not need to inspect the contents of the transaction to perform its operation. Ordering service creates blocks of transactions per channel, signs the blocks with its identity and broadcasts them to all peers using gossip messaging protocol. The main roles of this component are sorting received transactions, creating blocks and delivering them to all peers.

*3) Validation and Committing Phase:* All peers, both endorsing and committing peers on a channel receive blocks from ordering service. The peer will verify the order's signature on the block and then validate read set versions to decide whether transactions are valid or not. Each valid block will be committed to blockchain ledger while the write set in each transaction within the block will be updated to the world state which mainteins the latest value of states.

## B. Blockchain Structure

A blockchain is a growing list of records called blocks. Which are linked using cryptography. Each block contains a cryptographic hash of the previous block. The data in any given block cannot be altered retroactively without alteration of all subsequent blocks, which requires consensus of the network majority. On the other hand, blockchain can also be considered as a distributed ledger kept by peers. Important concepts of blockchain include:

*1) Chain:* Chain consists of a number of blocks which are linked together. The structure is similar to the list structure. Take Hyperledger Fabric for example, each block is identified with a hash value. Every block records the hash of the previous block called parent hash when they are generated. Therefore, with the parent hash recorded in the block header, this helps to ensure that each and every block is inextricably linked to its neighbour, leading to an immutable ledger.

*2) Block:* Block is used to record a set of transactions which are generated by invoking smart contracts. In Hyperledger Fabric, a block consists of three sections: block header, block data and block metadata. The block header records the basic information (e.g., block number, current block hash, previous block hash, etc.). The block data contains a list of transactions arranged in order. It is written when the block is created by the ordering service. The block metadata contains the time when the block was written, as well as the certificate, public key and signature of the block writer. Subsequently, the block committer also adds a valid/invalid indicator for every transaction. The whole block structure [26] is shown in Figure 3.

*3) Transaction:* A transaction represents an operation of the ledger which is generated by clients through invoking deployed chaincodes. It includes some important information (e.g. signature, read-write set and endorsements). The signature in a transaction is used for peers. In a Fabric network to validate transactions when peers receive a block broadcasted from the ordering service. The read-write set (RW-set) is the output of a smart contract. If the transaction is successfully validated, it will be applied to the ledger to update the world state. The Endorsements is a list of signed transaction responses from each required organization sufficient to satisfy the endorsement policy [5].

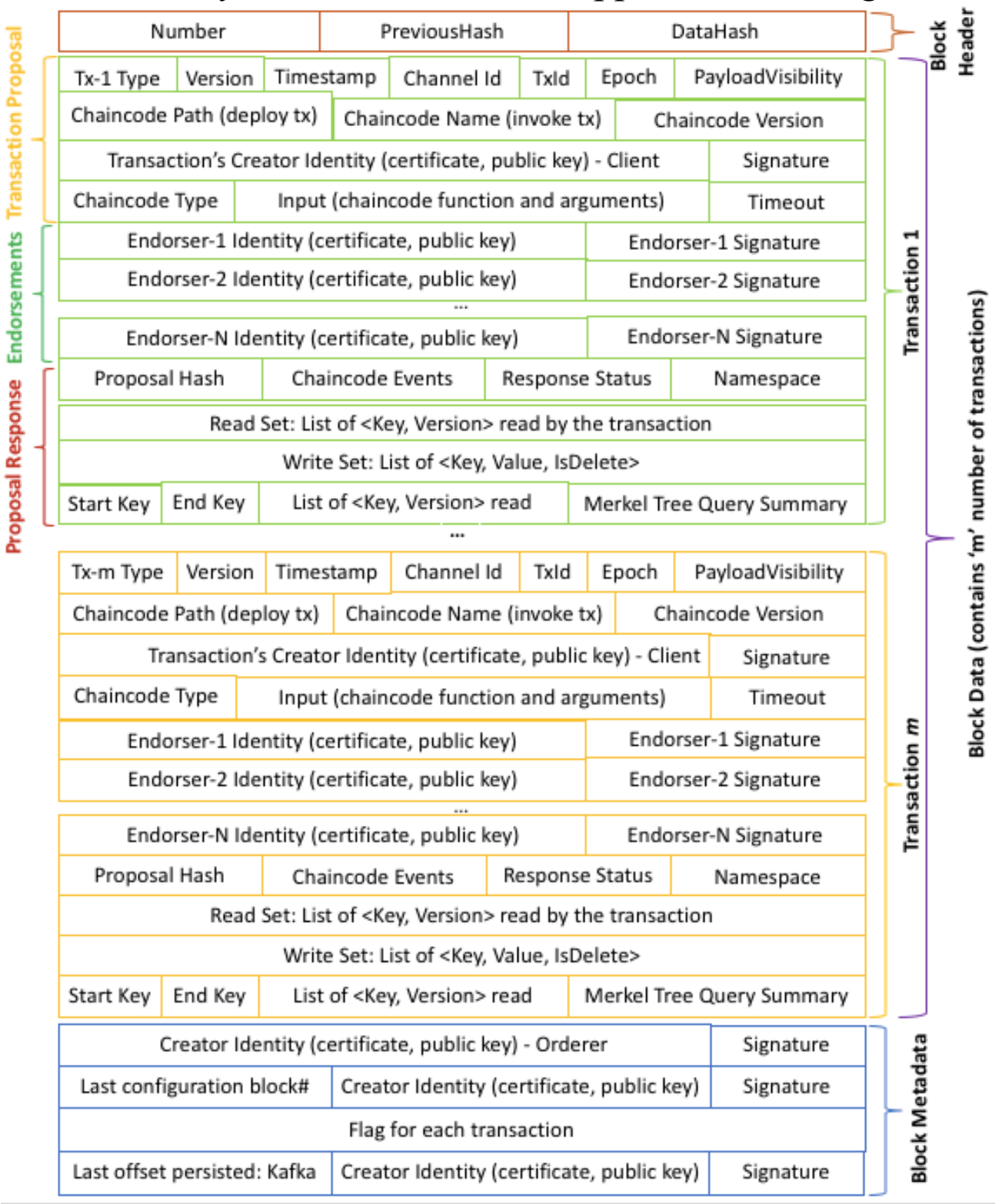

Fig. 3. Block structure in Hyperledger Fabric v1.0 [23]

## C. Data Storages

To better understand the data storages in Fabric, we briefly show the overview of all the data storages of Hyperledger Fabric version 1.1. The databases in Fabric is mainly comprised of four components in Figure 4.

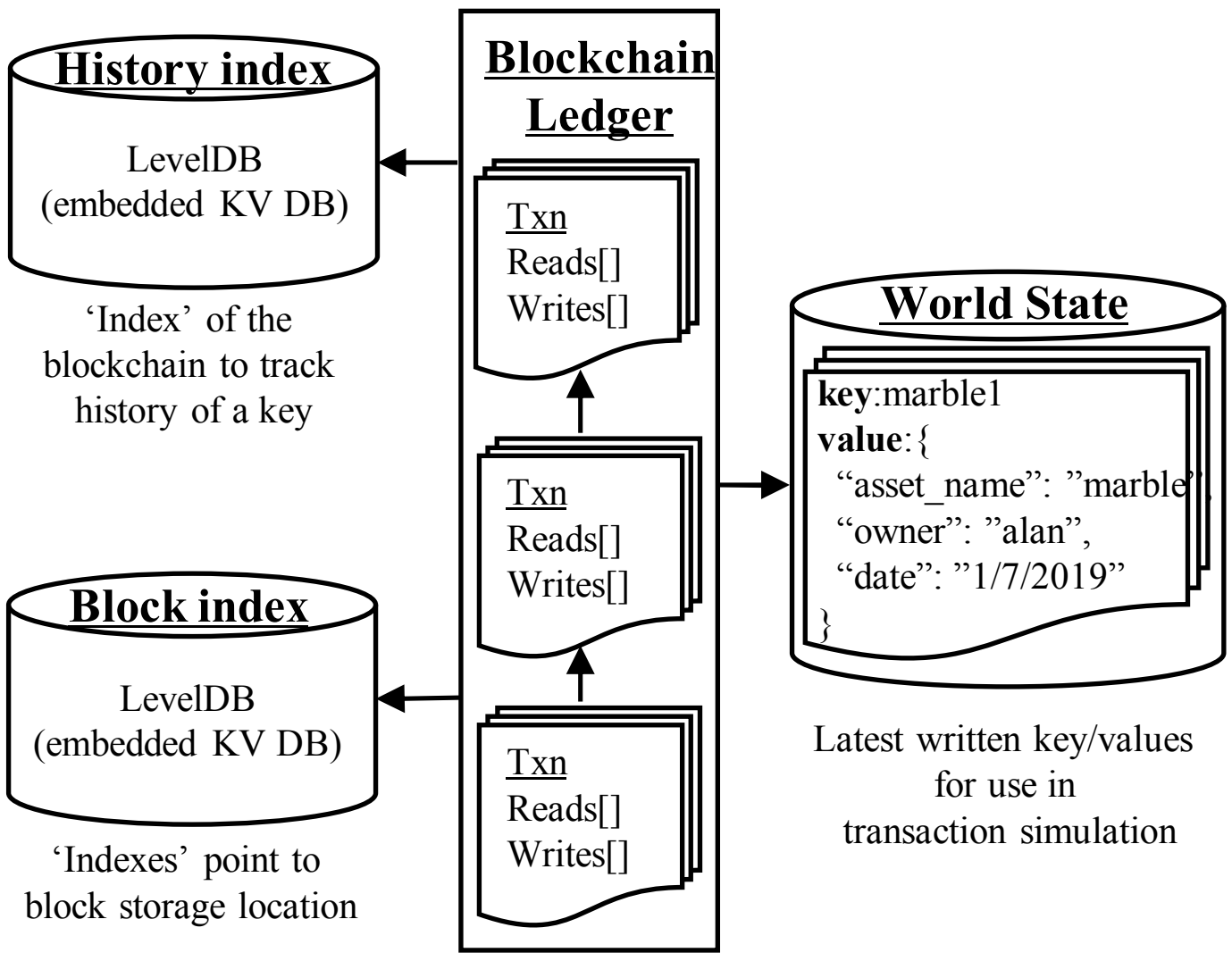

Fig. 4. Data storages in Hyperledger Fabric.

*1) Blockchain Ledger:* The blockchain ledger is comprised of a chain to store the immutable and sequenced record in blocks. It records all the changes that have resulted in the current world state. Transactions are collected inside blocks that are appended to the blockchain – enabling you to understand the history of changes that resulted in the current world state. It is immutable. The blockchain data structure is very different from the world state because it cannot be modified once written. There is one ledger per channel. Each peer maintains a copy of the ledger for each channel of which they are a member.

*2) World State:* A database that holds a cache of the current values of a set of ledger states. The world state makes it easy for a program to directly access the current value of a state rather than having to calculate it by traversing the entire transaction log. Ledger states are, by default, expressed as key-value pairs. They can be changed frequently as states can be created, updated and deleted. Currently, LevelDB [15] or CouchDB [27] can be choosed as the World State database in a Hyperledger Fabric system. The difference is that CouchDB supports rich queries of the states while LevelDB doesn't support. This database is designed for supporting the execution of chaincodes. The states in this database can be accessed through chaincode invocation.

*3) Block Index:* Whenever a block and the transactions within it are validated, this block and the transactions will be indexed into this database. It is used to facilitate users and developers to retrieve the specified block or transaction from the ledger. Users can use Fabric SDKs like Node SDK [28] to query blocks and transactions directly by using this index.

*4) History Index:* After a transaction is validated and committed into blockchain ledger, all the states in the Write Set of the transaction will be added into this index database is the form of <state, (block_num, tx_num)>. This enables users to track the operating history for a specific state. But there are some limitations to use this index database. First, it needs to start up this database when a Fabric network starts. Second, this history index only supports accessing from chaincodes. It is designed to meet the business needs of blockchain applications.

## IV. Problem Statement

Hyperledger Fabric is one of the most popular permissioned blockchain platform aims at enterprise-grade business applications. More and more enterprises begin to build applications based on Fabric. Fabric contains four data storages to meet business needs, some of them can be accessed through chaincode invocation only. Although Fabric Node SDK [28] provides some APIs to access the ledger, they are very simple. Users can only query a transaction by its id or get a block based on its block number.

In many scenarios, enterprise users need to do some complex business analysis on whole ledger data, or to make some fine-grained queries on transaction data. For example, In a food traceability project, querying information about a specific employee with the name 'David' (the query condition may like 'EmployeeInfo.EmployeeElement.Name="David"'). However, the interfaces provided by Fabric SDKs are far from satisfactory. From these needs, the three main problems of current systems are:

*1) Ledger Data Overview:* Currently, Fabric SDKs just provide APIs to retrieve block by its number or to query transaction by its id. However these simple APIs are not sufficient to satisfy practical data query requirements, e.g. retrieve transaction by defined creator / endorser / invoked function / channel id / timestamp etc.

*2) Tracking Operating History:* Users can access the history index through chaincode invocation to get the operating history of a certain state, but the precondition is that the history index service has started. Otherwise, there are no way to track a state's operating history. Meanwhile, there is a strong demand for this function under offline data analysis.

*3) Rich Query:* Although chaincode has provided function GetQueryResult() to perform a rich query against a state database (also known as world state database), but the precondition is to use CouchDB as the world state database. If using LevelDB as the world state database, this function will not work. In addition, The data structures of state values are not exposed. With the increase of states are stored, it's difficult for end users to perform rich queries on state values.

## V. Approach

We proposed a framework called Ledgerdata Refiner to solve the above problems. The main module of this framework is the data analysis middleware. It extracts and synchronizes data from blockchain ledger directly, and then parses data relationship to provide unified interfaces and UI for users to browse and query data conveniently. So, It works no matter what kind of database is used for world state and whether history index has been started or not. Figure 5 shows the structure of this framework.

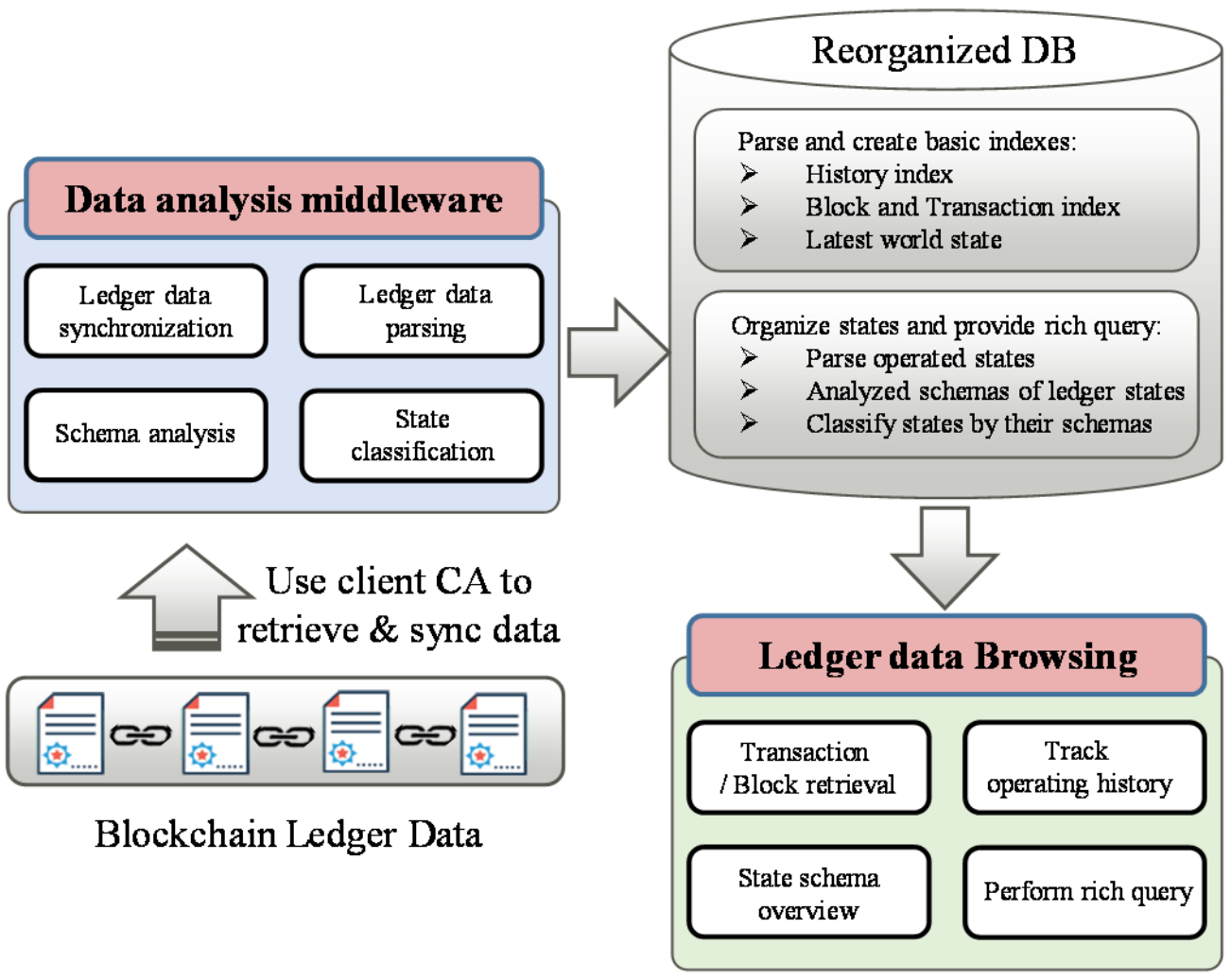

Fig. 5. The structure of Lerdgerdata Refiner

### A. Ledger Data Reorganization

*1) Ledger Synchronization:* The Ledgerdata Refiner could connect to any peer of a Fabric network as long as a client certification is provided. It can access and retrieve blockchain ledger data on this peer. When this tool retrieves blocks, the Fabric network can operate normally, it will not interfere the operation of the network. Some time will be taken for the initial synchronization of ledger. Once the synchronization has been finished, the maximum block height will be recorded and used to perform incremental synchronization next time. The ledger synchronization process is a regular task in Ledgerdata Refiner. Here we set it to synchronize every two seconds. LISTING 1 shows the synchronization strategy.

*2) Ledger Data Parsing:* Previous solutions expose some simple ways to retrieve blocks and transactions, such as block hash and transaction id. In fact, as we can see from Figure 3, block and transaction contain a variety of information, especially for the data structure of transaction. It records all the information from its creation to endorsement to submission, such as transaction creator, endorsers, invoked function and operated states et al. The detailed information we extracted from a block and a transaction are shown in Figure 6. In some practical applications, users may need to do some analytics and statistics on these information, which are not provided by current data access interfaces.

LISTING I. SYNCHRONIZATION STRATEGY

```
If The previous process is still running:
   exit()
Set max_block_height := retrieve it from the ledger data
Set current_block_height := recorded_block_height
If max_block_height > current_block_height:
   For block_num in (current_block_height,
   max_block_height]:
      retrieve block with block number "block_num"
      Set recorded_block_height := block_num
Parsing all the retrieved blocks
```

We parse out these information and reorganize them into a third-party database (e.g., PostgreSQL) to provide multiple query functions. For example, we can set different filters to retrieve transactions, such as creator, endorsers, time range and invoked function et al. In addition, we associate the operated states in each transaction with the transaction id and block number where they are located, in the form of <state, transaction_id, block_num, …>, to perform operating history tracking for any state.

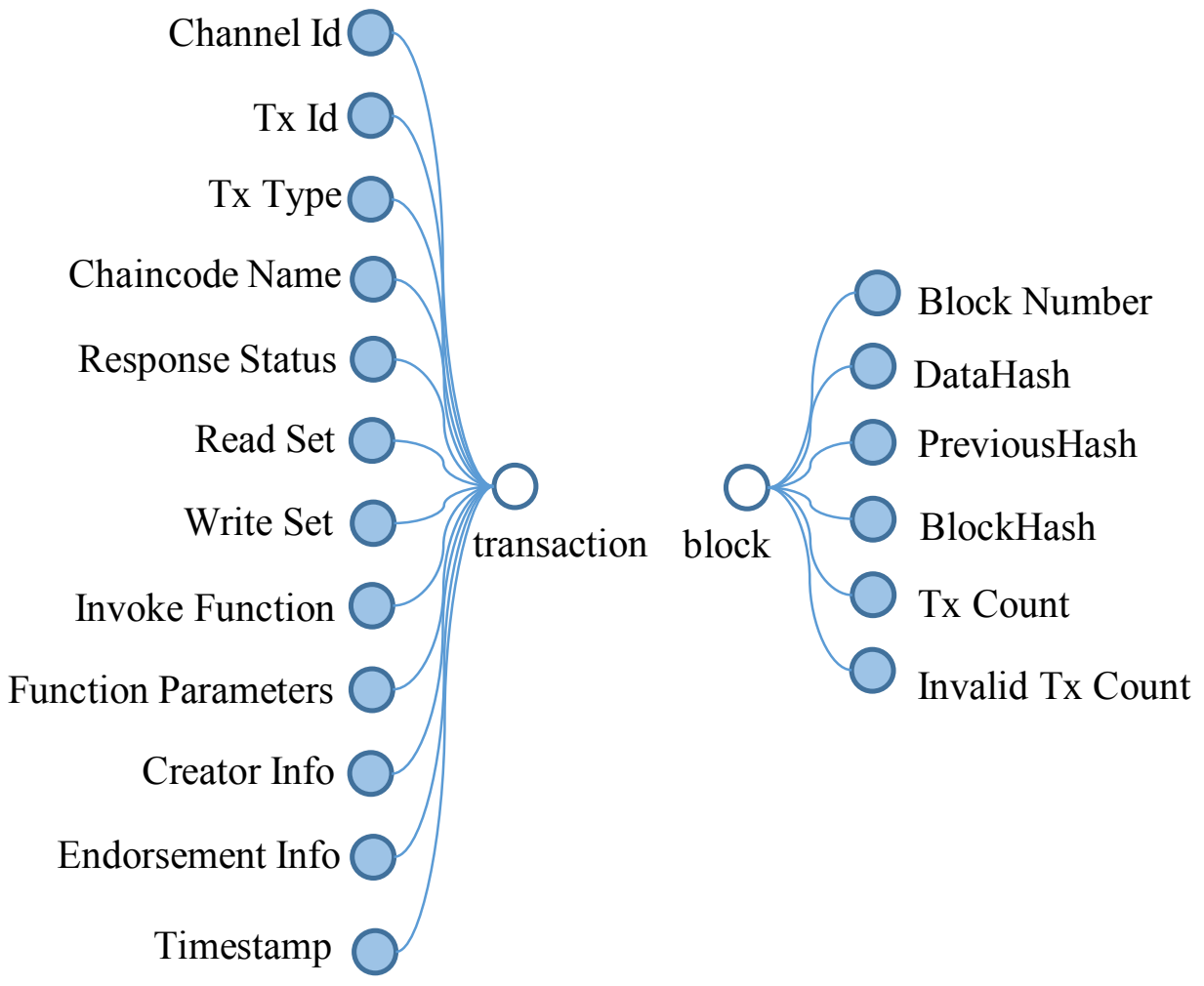

Fig. 6. The detailed information extracted from block and transaction

## B. *State Schema Analysis*

The table structure of a traditional relational database indicates the data structure of the stored records. But in Fabric, the states is stored in the form of <key, value> in the ledger. In most cases the value is in json format. It is difficult for developers and users to identify the json schema of the stored states as time goes on. It is also inconvenient to retrieve json content according to query requirements. So in this section, we proposed a method to display the schemas of the stored states and use them to assist rich query.

*1) Schema Extraction:* In the process of synchronizing blocks, we perform schema extraction for each state in the write set of every transaction and mark each state with its extracted schema. These extracted schemas are used to classify the states dynamically during the process of incremental synchronization of ledger data and are also used to assist users in rich query of state values. Figure 7 shows an example of schema extration.

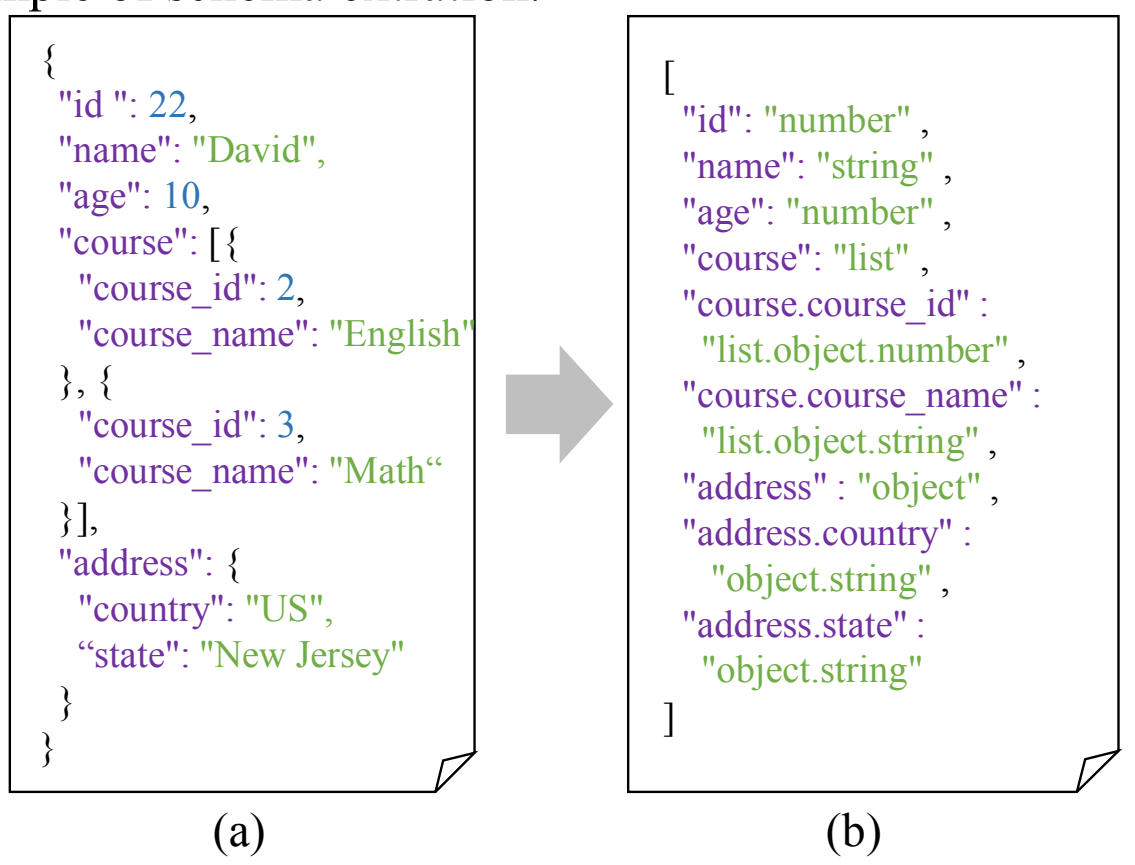

Fig. 7. An example of schema extraction. (a) A state in json format. (b) The extracted schema with hierarchical relationships and data types

*2) Schema Similarity Comparison:* We maintain a schema table to store the parsed schemas. In the process of incremental synchronization of ledger data, for the extracted schema of a certain state, we compare it with all the schemas in this table we update the table according to the schema comparison result by keeping the correspondence of schema id between the two tables (as Figure 8 shown). Four schema comparison results are depecited as follows:

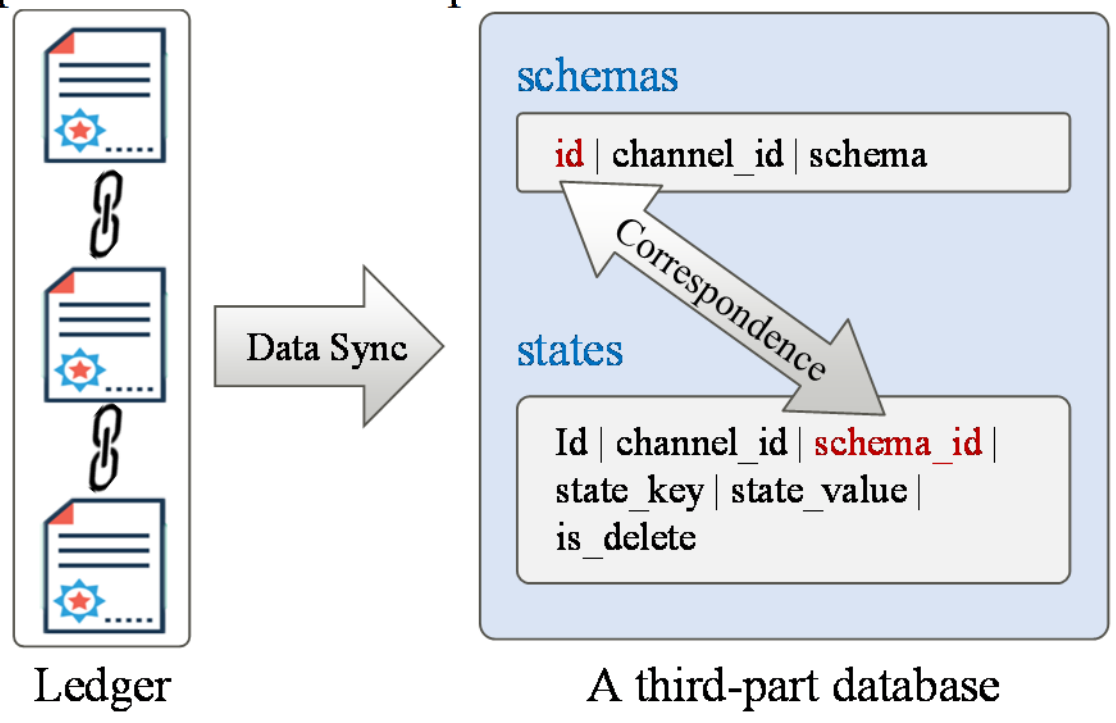

Fig. 8. Using extracted schemas to classify ledger states

- **Equal:** If the extracted schema is equal with one of the schemas which stored in schema table, the existing schema id will be used directly to represent the new coming state.
- **Contain:** If the content of an existing schema in the schema table is a partial of that of the new extracted one, the new schema content will be update into the table to replace the existing one. However, the existing schema id keeps unchanged.
- **Be Contained:** Suppose that the content of the extracted schema Is a partial of that of an existing

schema in the schema table, the existing schema id will be used to represent the new coming state directly.

- **Different:** If the extracted schema is different with any schema in the schema table, we will insert the new schema into the schema table and using its id to represent this state.

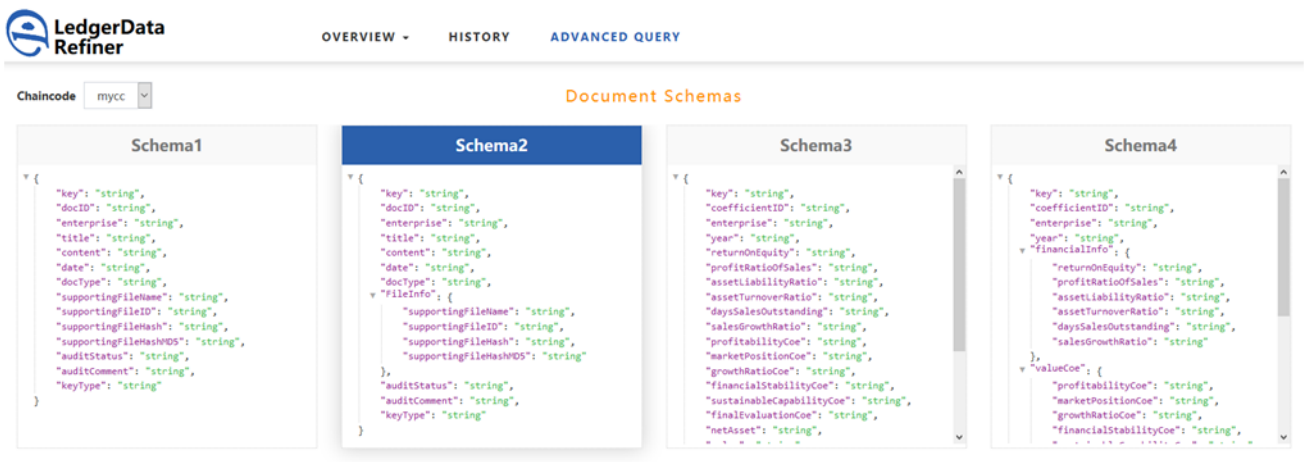

(a) Select a schema to perform rich query

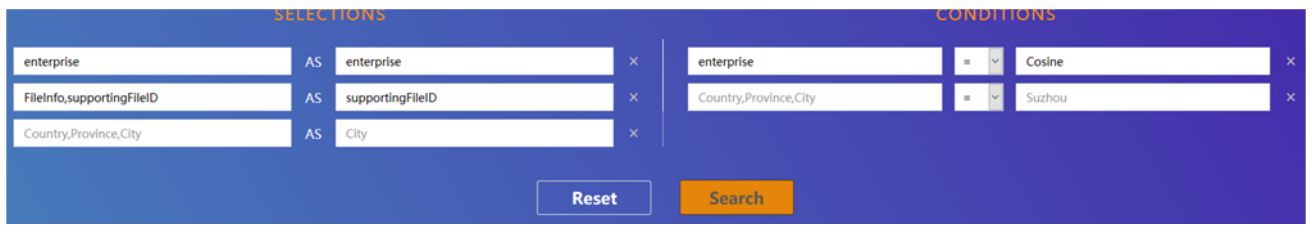

(b) Set query fields and conditions based on the selected schema

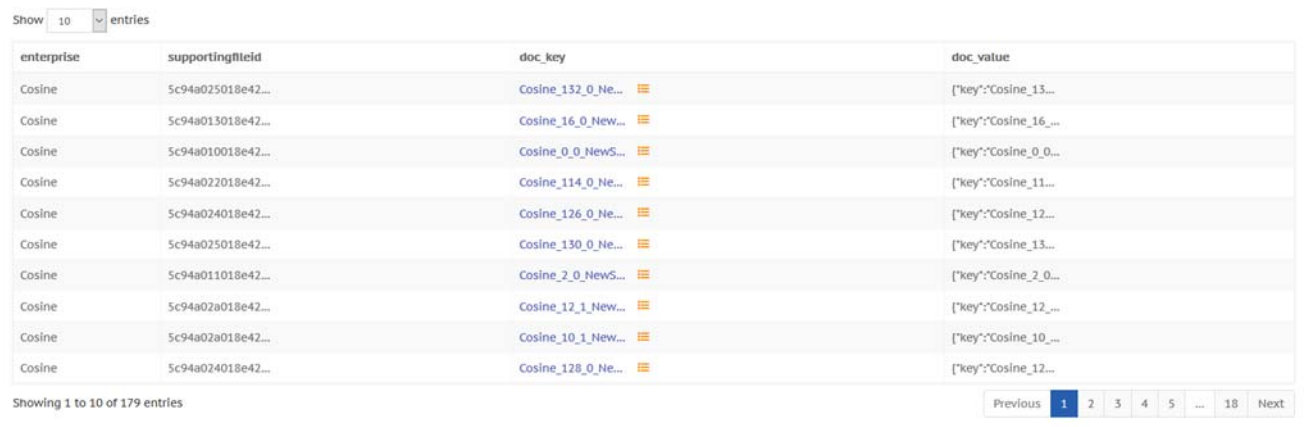

(c) Query results

Fig. 9. An example of performing a rich query on Ledgerdata Refiner

In order to display an overall view of schemas, we choose PostgreSQL as the third-party relational database. Two tables are created to support rich queries against ledger states. One is used for storing the analyzed schemas. The other is used for storing states with json as their data type. This enables users to view state schemas and assist them to perform rich queries on state values. We compare our platform with other tools as shown in Table I. Figure 9 shows an example of rich query with the knowledge of analyzed state schemas as precondition.

## VI. Environment and Experiment

In this section, we test the Ledgerdata Refiner platform based on a real application scenario. We built a Fabric test network and submitted 100,000 transactions to test the performance of our framework.

### A. Experimental Background

Small and medium-size enterprises (SMEs) often face difficulties in financing because they do not meet the requirements for IPO. It is also difficult for investors to dig out high-growth companies from thousands of companies. In order to balance the needs between investors and SMEs, we developed an information disclosure platform based on Fabric for SME alliance. This application is mainly used to record various business reports and other disclosed information and maintain the immutability of disclosed information by using blockchain technologies.

The first phase of this application consists of two organizations. One is council, and the members are elected from SMEs. The other is enterprise which represents SMEs. The council organization is responsible for reviewing members of enterprise organization and the information disclosed by them. The members in enterprise organization can disclose their business reports and other information about themselves. Potential investors can browse the disclosed information to find companies of high quality. The main modules of this platform is shown in Figure 10.

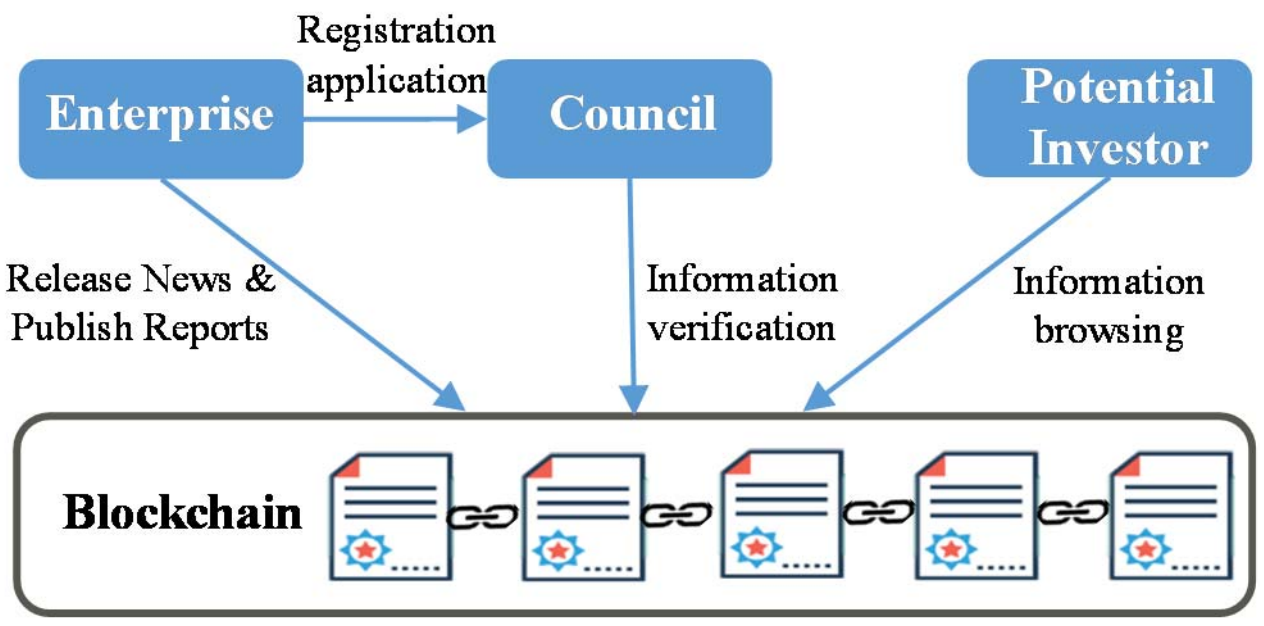

Fig. 10. Function modules of Information Disclosure Platform

### B. Experimental Setup

Figure 11 shows the experiment setup we used in our experiments. We deployed the Information Disclose Platform for testing. The blockchain consortium of this setup contains two organizations Org1 and Org2, the Org1 represents the organization of council and the Org2 represents the organization of enterprise. Each contributed two peers to this blockchain network. The endorsement policy on transactions is set to include signatures from at least one peer from any organization to successfully commit on the blockchain. One channel is set up between both organizations. All the chaincodes were deployed on this channel. The Ordering Service Node (OSN) is running under the solo model. All these peers are running on a single virtual machine with 4vCPUs (4 cores at 3.0 GHz with hyper threading) and 6GB RAM, where the operating system is the Ubuntu 16.04 LTS. After we finished importing transactions into this Fabric network, we run the data synchronization and processing programs on this virtual machine.

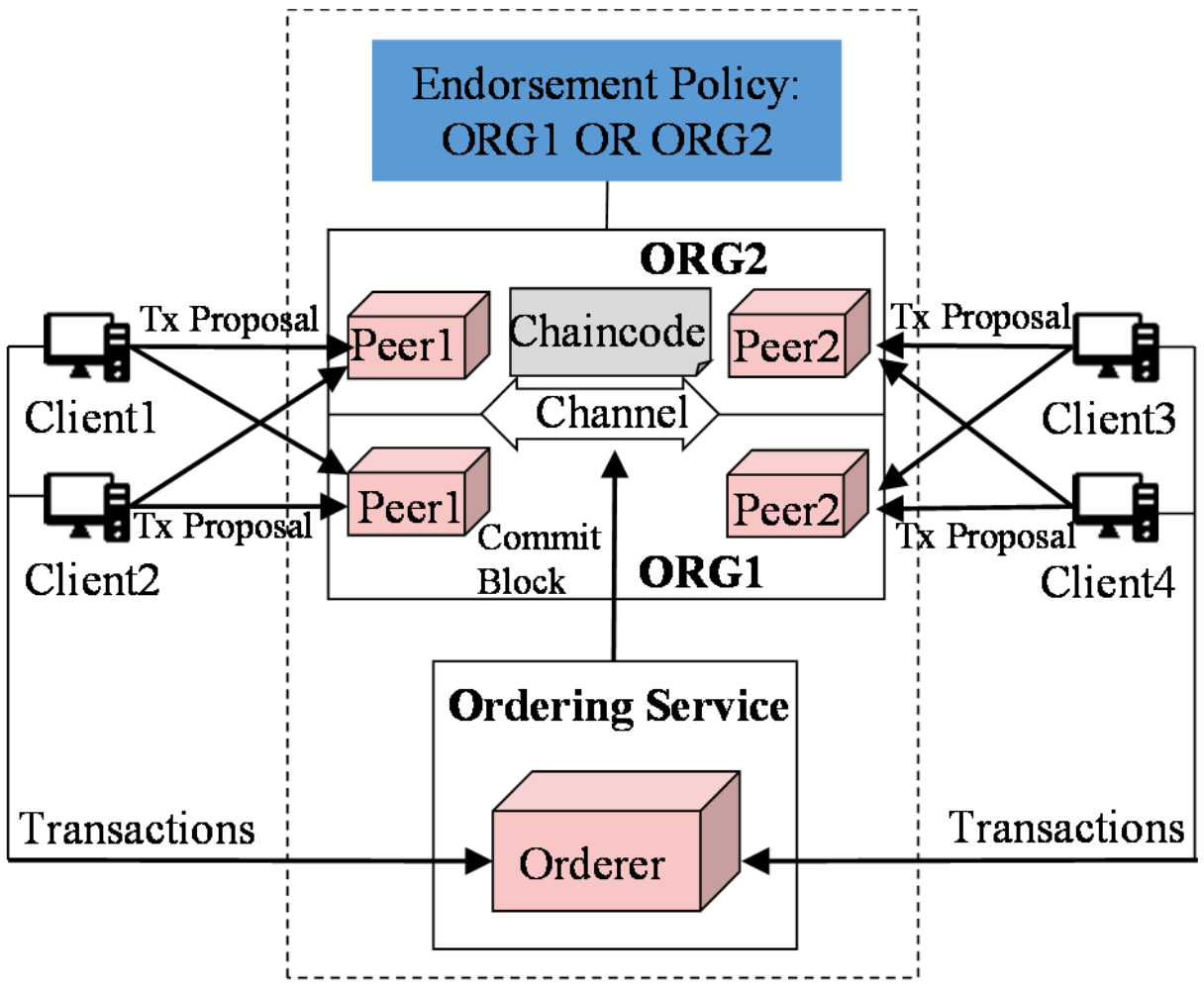

Fig. 11. Fabric construction used in experiments

### C. Evaluation

The information disclosure platform currently has four types reports with json format. Two types of the json reports are one-level structure. The other two are two-level structure. Table II shows the detailed number of properties of each level.

TABLE I. COMPARISON WITH OTHER TOOLS

| Category | Description | Ledgerdata Refiner | Hyperledger Explorer | Fabric SDK | Chaincode |
|---|---|---|---|---|---|
| Usage of data | How to use the ledger data | Offline | Offline | Online | Online |
| Visualization | Visualization of ledger data | √ | √ | | |
| Ledgerdata overview | Statistics of block and tx | √ | √ | | |
| | Block browsing | √ | √ | √ | |
| | Transaction browsing | √ | √ | √ | |
| | Basic search of block and tx | √ | | √ | |
| History query | History query of any state | √ | | | √ |
| Advanced query | Schema analysis of states | √ | | | |
| | Perform a 'rich' query of states | √ | | | √ |

TABLE II. OVERVIEW OF FOUR SCHEMA

| State Structure | Number of Levels | Property Number | |
|---|---|---|---|
| Schema-A | 1 | 14 | |
| Schema-B | 2 | 1st level | 11 |
| | | 2nd level | 4 |
| Schema-C | 1 | 24 | |
| Schema-D | 2 | 1st level | 12 |
| | | 2nd level | 15 |

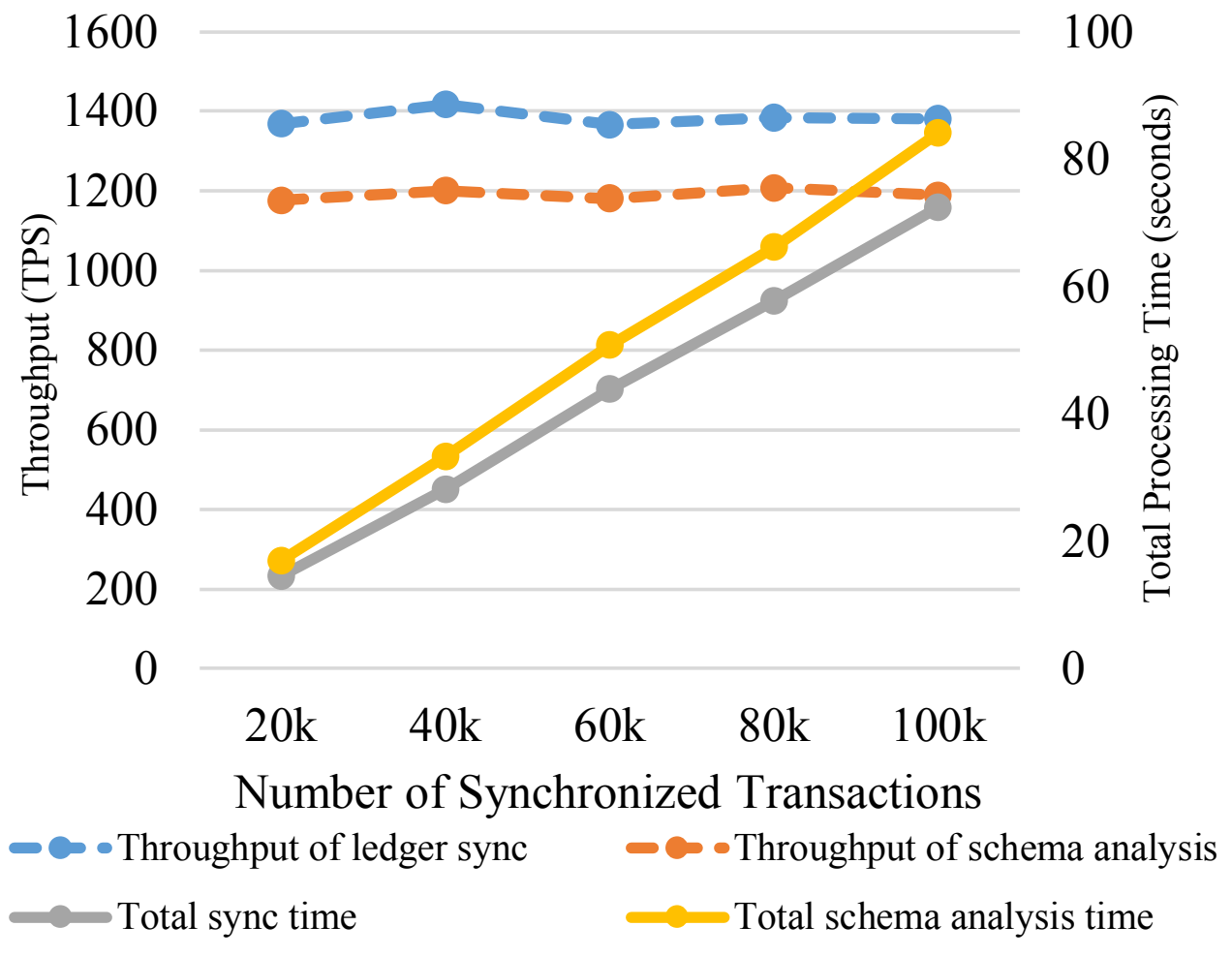

Fig. 12. Performance of Ledgerdata Refiner

In order to better test the data analysis performance of our Ledgerdata Refiner platform, we created 100,000 reports based on four json schemas and imported them to the mentioned experimental Fabric network by submitting transactions. The Ledgerdata Refiner has two main processing processes. One for ledger data synchronizing and parsing, the other for schema analysis for states. They can execute in parallel. We use Ledgerdata Refiner to process the 100,000 transactions and record the performance of the two main processing processes. The performance is shown in Figure 12. From this figure we can see the throughput of ledger data synchronization and parsing process is about 1,400 TPS (transactions per second). The throughput of schema analysis process is about 1,200 TPS. As can be seen from the figure, the processing time of these two processes is linear with the number of transactions. They are relatively stable. According to the experimental environment and the throughput that the current Fabric network can withstand [6], our framework can adequately satisfy it. In addition, this Ledgerdata Refiner is able to handle more complex transactions like financial data for the ledger data is in a uniform format in Fabric.

## VII. CONCLUSION

The main contribution of our work is to construct a ledger data query platform called Ledgerdata Refiner for permissioned blockchain system such as Hyperledger Fabric. Its key component is ledger data analysis middleware, which extracts and synchronizes ledger data, and then parses the relationship among them. With this middleware, besides query blocks and transactions, we also provide enriched data view for end users, including schema overview and customized fine-grained query on ledger states. In the future, we hope to make further research on querying and analyzing on Hyperledger Fabric ledger data, to provide sufficient and efficient interfaces for end users.

## ACKNOWLEDGMENT

We sincerely thank the members of Fujitsu Laboratories Ltd. for supporting our research and providing clear guidance in this research area. Besides, we thank our colleagues for discussing with us together and designing the UI of the Ledgerdata Refiner.